\documentstyle[aps,epsf,pra%
,twocolumn%
]{revtex}
\newcounter{US:Fv2}
\begin{document}
\draft
\title{On the driven Frenkel-Kontorova model: II. Chaotic sliding and 
nonequilibrium melting and freezing}
\author{Torsten Strunz and Franz-Josef Elmer}
\address{Institut f\"ur Physik, Universit\"at
   Basel, CH-4056 Basel, Switzerland}
\date{{\tt submitted to Phys. Rev. E, September, 1997, revised Jan. 98}}
\maketitle
\begin{abstract}
The dynamical behavior of a weakly damped harmonic chain in a
spatially periodic potential (Frenkel-Kontorova model) under the
subject of an external force is investigated. We show that the chain
can be in a spatio-temporally chaotic state called fluid-sliding
state. This is proven by calculating correlation functions and
Lyapunov spectra. An effective temperature is attributed to the
fluid-sliding state. Even though the velocity fluctuations are
Gaussian distributed, the fluid-sliding state is clearly not in
equilibrium because the equipartition theorem is violated.  We also
study the transition between frozen states (stationary solutions) and
molten states (fluid-sliding states).  The transition is similar to a
first-order phase transition, and it shows hysteresis. The
depinning-pinning transition (freezing) is a nucleation process. The
frozen state contains usually two domains of different particle
densities.  The pinning-depinning transition (melting) is caused by
saddle-node bifurcations of the stationary states. It depends on the
history.  Melting is accompanied by precursors, called micro-slips,
which reconfigurate the chain locally. Even though we investigate the
dynamics at zero temperature, the behavior of the Frenkel-Kontorova
model is qualitatively similar to the behavior of similar models at
nonzero temperature.
\end{abstract}

\narrowtext

\section{Introduction}

Systems with many degrees of freedom which are pinned in some
external potential are very common in condensed matter. Examples are
fluid-fluid interfaces in porous media \cite{rub89,he92}, 
flux-lattices in type-II superconductors \cite{bla94}, and charge
density waves \cite{gru88} to mention only a few. Also dry friction
(i.e., solid-solid friction) belongs to this class of systems because
the asperities of the surfaces interlock.

A common feature of all these systems is a strongly nonlinear
mobility. If one applies some field or force $F$ on the system, the
mobility is zero below some usually well-defined threshold $F_c$.
Above this threshold the mobility is nonzero. In general it is some
nonlinear function of the applied force $F$.  The transition from a
pinned system with zero mobility to a depinned one with some finite,
nonzero mobility is called the {\em pinning-depinning transition\/}.
It can be understood as a kind of ``melting'' which happens far from
thermal equilibrium. The process is a typical nonequilibrium one
because of two reasons. First, there is no ground state for $F\neq 0$
and the pinned system has to be in some metastable state. Due to
thermal fluctuations the system can overcome the barrier of the
metastable state and move into another metastable state with less
energy. This phenomenon leads to {\em creeping\/} with a very low
mobility.  Second, beyond the pinning-depinning transition energy
flows through the system at a constant rate which is given by the
mobility times $F^2$. This flow is usually not small. Thus it cannot
be deduced from linear response theory which works only near thermal
equilibrium. The mobility of the sliding state strongly depends on
the kind of energy dissipation.

The inverse process of this nonequilibrium melting is the {\em
depinning-pinning transition\/} which is a kind of nonequilibrium
``freezing''. Both kind of transitions do not have to occur at the
same value of the applied force $F$. The behavior depends strongly on
whether the degrees of freedom (i.e., flux lines, atoms etc.) have
inertia or inertia is negligible compared to dissipative forces. In
the case of strong dissipation the motion is overdamped. The
pinning-depinning transition is in most cases of second order and
indistinguishable from the depinning-pinning transition. Typical
examples for such systems are flux lines in type-II superconductors
and charge-density waves.  If the motion is underdamped, hysteresis
is possible because the inertia can overcome a pinning center. This
is intuitively clear if one imagines the simplest model system of
this kind, namely a particle in a spatially periodic potential
\cite{ris84}.

Another important aspect of the collective behavior of pinned systems
is whether the potential caused by the pinning centers is regular or
irregular (quenched randomness). Often the pinning landscape is
random. This case together with a purely dissipative diffusion-like
dynamics has been studied extensively in the literature
\cite{nat92,nar93}. 

The aim of this paper is to study the opposite case in a fairly simple
model, namely, the Frenkel-Kontorova (FK) model \cite{kon38}. There
is no quenched randomness. All pinning
centers are identical forming a regular array. Furthermore, all
pinned objects are identical and have a mass.
The damping is assumed to be weak. 
We will see that weak damping is responsible for randomness that 
is caused by chaotic motion.
Important physical applications of the FK model are arrays of
identical Josephson junctions \cite{wat96} and adsorbate layers on
clean crystal surfaces \cite{bra97a}. 

In this paper we consider the one-dimensional
FK model. The equation of motion (in dimensionless units) reads
\begin{equation}
  \ddot x_j+\gamma\dot x_j=x_{j-1}+x_{j+1}-2x_j-b\sin x_j+F,
  \label{INT:eqm}
\end{equation}
where $x_j$ is the position of particle $j$, $\gamma$ is the damping
constant, $b$ is the strength of the external potential, and $F$ is
the external force. The time derivative is denoted by a dot. In order
to avoid effects due to the boundary layers we choose periodic
boundary conditions, i.e.,
\begin{equation}
  x_{j+N}=x_j+2\pi M,
  \label{INT:bc}
\end{equation}
where $N$ is the number of particles and $M$ is an arbitrary integer.
The periodic boundary conditions fixes the average particle distance
$a$ to $a=2\pi M/N$. Because of symmetry $a$ can be restricted to 
interval $[0,\pi]$ without loss of generality..

Together with the previous paper \cite{str97} in which we have
already investigated periodic and quasiperiodic solutions, the aim is
to give a detailed investigation of the dynamical behavior in the
weakly damped case for long chains (i.e., $N>100$).

\begin{figure}
\epsfxsize=80mm\epsffile{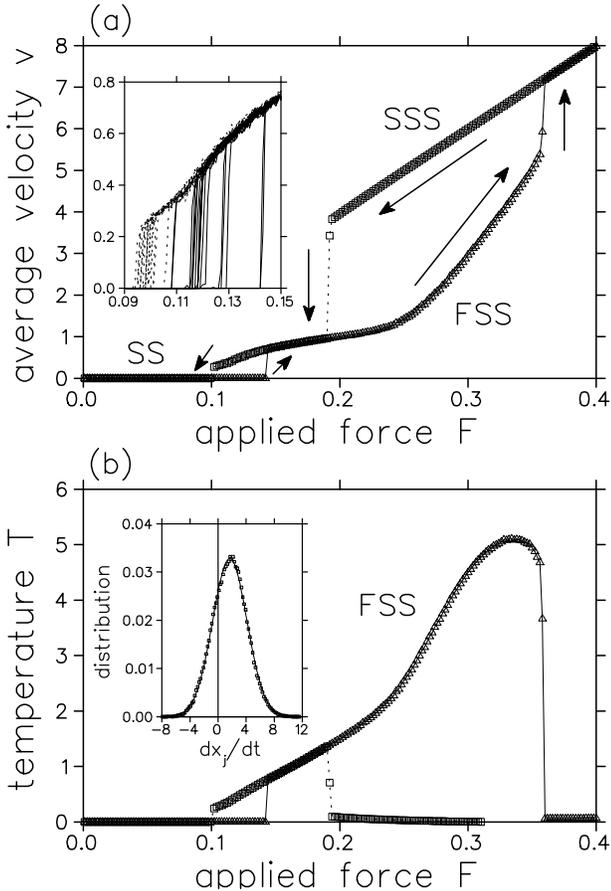}\vspace{2mm}
\caption[Velocity-force characteristic and Temperature]{
\protect\label{f.vF.chaos} The velocity-force characteristic and the
effective temperature of the fluid-sliding state.  The different
branches belong to stationary states (SS), fluid-sliding states
(FSS), and solid-sliding states (SSS). In the simulations the applied
force $F$ was decreased (squares and dotted lines) or increased
(triangles and solid lines) with a constant rate ($|\dot
F|=10^{-7}$). The velocity at each data point is the average over a
time interval of $10^4$ time units. The upper inset shows 20
hysteresis loops between SS and FSS from a simulation where $F$ was
moved in the interval $[0.09,0.15]$ forward and backward at a rate of
$|\dot F|=10^{-6}$.  The lower inset shows the particle velocity
distribution for $F=0.26$. The solid line shows the fit of the data
point with a Gaussian.  The parameters are $N=233$,
$M=89$, $b=2$, and $\gamma=0.05$.}
\end{figure}

In this paper we deal with spatio-temporal chaotic solutions (called
{\em fluid-sliding states\/}) and the transition between these
solutions and the stationary states. The typical behavior is
summarized in figure~\ref{f.vF.chaos}. Fig.~\ref{f.vF.chaos}(a) shows
the velocity-force characteristic. We see hysteresis loops between
three different branches which belong to different types of solutions.
The states with the largest average sliding velocities $v$ are the
{\em solid-sliding states\/} which are characterized by a chain with
nearly no internal vibrations. These states have the maximum possible
mobility, i.e., $1/\gamma$. The solid-sliding state becomes unstable
due to first-order parametric resonance if $v$ is below some critical 
value\cite{str97}. 

The second type of sliding state is the fluid-sliding state. In
general one can summarize all sliding states having not the maximum
mobility into this category. But strictly speaking the name makes
only sense if these states are spatio-temporally chaotic. For larger
values of the damping constant this is not the case as we have seen
in the previous paper. The chaotic vibration in the fluid-sliding
state can be characterized by an {\em effective temperature\/}. The
temperature of the fluid-sliding states of Fig.~\ref{f.vF.chaos}(a)
is shown in figure~\ref{f.vF.chaos}(b). Even though the distribution
of the particle velocity $\dot x_j$ is gaussian [see inset of
Fig.~\ref{f.vF.chaos}(b)] we will show that the fluid-sliding state
is a {\em nonequilibrium\/} state. A very specific test to show this
is the violation of the equipartition theorem for the phonon modes
(see Sec.~\ref{STC}).

The third type of states are the stationary ones. Their mobility is
zero. In order to model the creeping due to thermal activation one
has to add a white-noise term to the equation of motion
(\ref{INT:eqm}). We have not done this because the qualitative
behavior does not change very much as long as the thermal energy is
much smaller than the amplitude of the external potential. This is
confirmed in numerical simulations of similar models
\cite{bra97a,per93,gra96}. For example, the hysteresis seen in
Fig.~\ref{f.vF.chaos}(a) still exist for nonzero but small
temperatures \cite{bra97a,bra97b}. 
For this behavior it seems to be important that the
system has many degrees of freedom because in the case of $N=1$ the
hysteresis disappears even for infinitesimal small noise amplitude
\cite{ris84}.

Fig.~\ref{f.vF.chaos}(b) clearly shows that nonequilibrium melting and
freezing is accompanied by an abrupt change of the temperature of the
chain. The transition is like a first-order one in thermal
equilibrium but the pinning-depinning transition point is larger than
the depinning-pinning transition point. Thus hysteresis occurs. The
transition points fluctuate [see inset of Fig.~\ref{f.vF.chaos}(a)],
especially the pinning-depinning transition point.

The paper is organized as follows: In section~\ref{CS} we investigate
in detail the fluid-sliding state. We show that it is indeed
spatio-temporally chaotic. For chains with $a/2\pi$ near an integer
value we found a pronounced transition from a kink-dominated
sliding state and the fluid-sliding state. This transition is relatively
sharp even though there is no hysteresis. But it becomes hysteretic
for small $N$. The depinning-pinning transition and the
pinning-depinning transition are discussed in section~\ref{MF}. We show
that nonequilibrium freezing is similar to ordinary freezing whereas
melting is clearly different. The pinning-depinning transition point
depends on the stationary state. Furthermore local rearrangements
(micro-slips) of the chain may occur before the transition. In
section~\ref{CON} we compare our results with results of similar
models.

\begin{figure}
\epsfxsize=80mm\epsffile{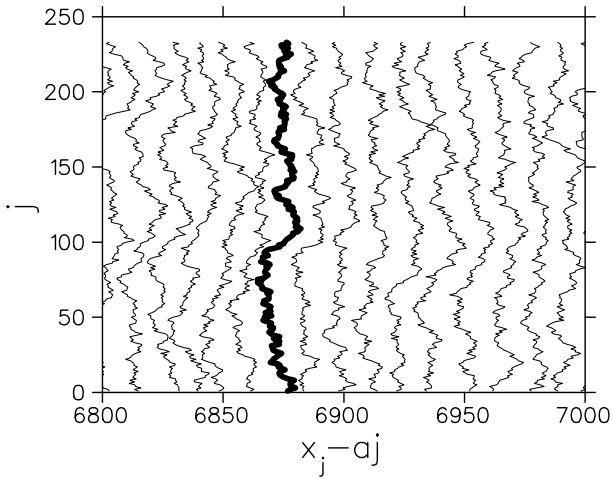}\vspace{2mm}
\caption[Spatio-temporal chaos]{\protect\label{f.chaos}
An example for spatio-temporal chaos. Each solid line is a snapshot
of the system. The time interval between two successive snapshots is
$\delta t=4\pi/v$ ($v$ is the average sliding velocity). A particular
snapshot is highlighted. The parameters are
$N=233$, $M=89$, $b=2$, $\gamma=0.05$, and $F=0.14$.
}
\end{figure}

\section{\protect\label{CS}Chaotic sliding}

Decreasing the damping constant $\gamma$ increases the complexity of
the sliding state from periodic motion via quasi-periodic motion
(which is usually spatially chaotic, see the preceding paper) to
spatio-temporal chaos. An example of the latter one is shown
in figure~\ref{f.chaos}. 

The aim of this section is to investigate and to characterize the
chaotic-sliding state which we call the {\em fluid-sliding state\/}.
First of all we see in Fig.~\ref{f.vF.chaos}(a) that the
velocity-force characteristic of this state is nearly structureless.
This has to be compared with the case of periodic and quasiperiodic
motion where a multitude of hysteresis loops appear (see the
preceding paper). Here there occur only hysteresis loops between the
solid-sliding state (where the particles are shaken so fast that they
nearly do not feel the external potential), the fluid-sliding state
and the stationary states. 

\begin{figure}
\epsfxsize=80mm\epsffile{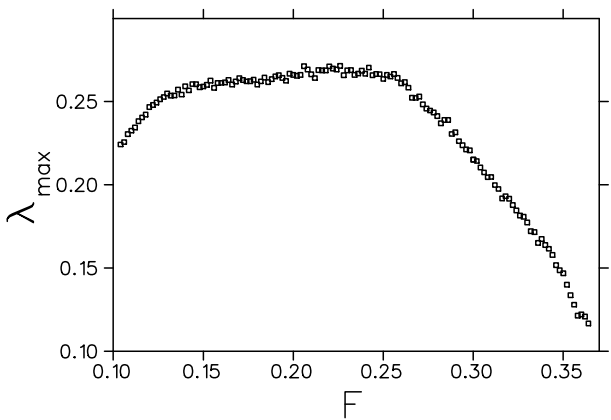}\vspace{2mm}
\caption[Maximum Lyapunov exponent]{\protect\label{f.lmax}
The maximum Lyapunov exponent of the fluid-sliding state 
as a function of the applied force $F$.
The parameters are $N=144$, $M=55$, $b=2$, and $\gamma=0.05$.
}
\end{figure}

\subsection{\protect\label{STC}Spatio-temporal chaos}

Fig~\ref{f.chaos} is of course not a proof that the chain slides
chaotically. It is well-known that chaotic motion is characterized by
the sensitivity on the initial conditions. It is measured by the
largest Lyapunov exponent $\lambda_{\rm max}$, which is the rate of
divergence (or convergence, if it is negative) of trajectories in
phase space that start out infinitely close to each other
\cite{cro93}. Figure~\ref{f.lmax} shows that the fluid-sliding states
in Figs.~\ref{f.chaos} and \ref{f.vF.chaos} are indeed temporally
chaotic.

\begin{figure}
\epsfxsize=80mm\epsffile{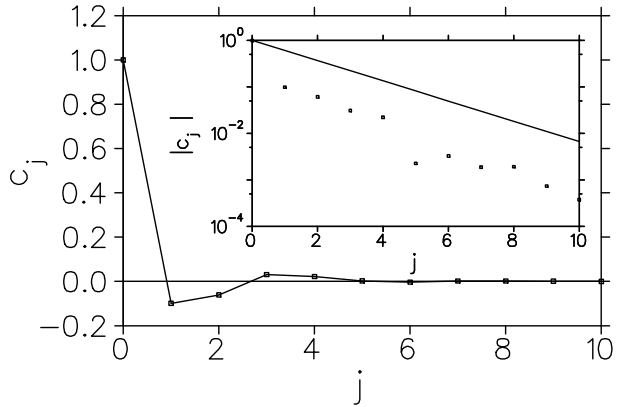}\vspace{2mm}
\caption[Velocity correlation function]{\protect\label{f.cf}
The normalized velocity correlation function $c_j$ of the
fluid-sliding state. To guide the eye the numerical results (denoted
by squares) are connected by a solid line. The inset shows a
logarithmic plot of $|c_j|$.
The straight line is the function $\exp(-j/2)$. The parameters are
the same as in Fig.~\ref{f.chaos}.
}
\end{figure}

But is it also spatially chaotic? In order to answer this question we
have calculated the normalized velocity correlation function $C_j$
defined by
\begin{equation}
  C_j\equiv\frac{\langle\!\langle\dot x_l\dot x_{l+j}\rangle\!\rangle-
   \langle\!\langle\dot x_l\rangle\!\rangle^2}{\langle\!\langle\dot 
   x_l^2\rangle\!\rangle-\langle\!\langle\dot x_l\rangle\!\rangle^2},
  \label{STC:cf}
\end{equation}
where 
\begin{equation}
 \langle\!\langle f_j\rangle\!\rangle=\lim_{\tau\to\infty}\frac{1}
 {\tau}\int_0^\tau\frac{1}{N}\sum_{j=1}^Nf_j(t)\,dt.
  \label{STC:d.av}
\end{equation}
For the same parameters as in Fig.~\ref{f.chaos} the result is shown
in figure~\ref{f.cf}. One clearly sees that $C_j$ is a rapidly
decaying oscillatory function. The envelope seems to be proportional
to $\exp(-j/\xi)$ with a correlation length $\xi\approx 2$. Because
of $N\gg\xi$ and $\lambda_{\rm max}>0$ the fluid-sliding state is
spatio-temporally chaotic.

\begin{figure}
\epsfxsize=80mm\epsffile{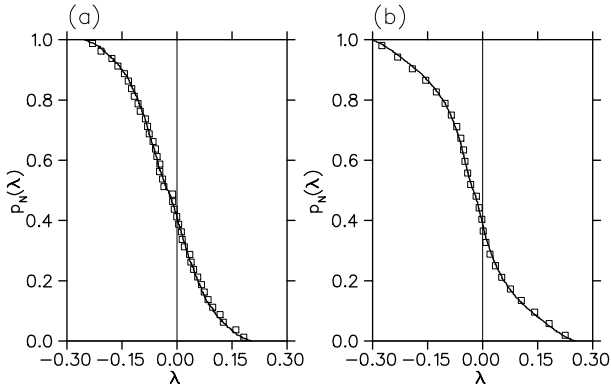}\vspace{2mm}
\caption[Lyapunov spectrum]{\protect\label{f.lspect}
Lyapunov spectra for (a) the most commensurate case (i.e. $a=0$) and
(b) the most incommensurate case (i.e. $a/2\pi\to (3-\sqrt{5})/2=2-$
golden mean). The spectra for two different system sizes are shown.
Squares and solid lines denote (a) $N=20$, (b) $N=13$, and (a)
$N=200$, (b) $N=233$. The other parameters are $b=2$, $\gamma=0.05$,
and (a) $F=0.13$, (b) $F=0.3$.
}
\end{figure}

A very strong criterion for spatio-temporal chaos is that the number
of positive Lyapunov exponents is proportional to $N$ for large $N$.
We have calculated the Lyapunov spectrum with the method described in
Ref.~\onlinecite{eck85} for various values of $N$. 
Figure~\ref{f.lspect} shows the cumulative density $p_N(\lambda)$ of
Lyapunov exponents, i.e., the probability to find a Lyapunov exponent
larger than $\lambda$. The result is typical for spatio-temporal
chaos \cite{cro93}. In the thermodynamic limit (i.e., $N\to\infty$)
the sequence of cumulative densities $p_N$ converges uniformly to
$p_\infty$. Thus, for large $N$ the number of positive Lyapunov
exponents is indeed proportional to $N$. Fig.~\ref{f.lspect} 
shows clearly that the spatio-temporally chaotic nature of the fluid-sliding state
does {\em not\/} depend on the commensurability of the chain. 

Because the fluid-sliding state is spatio-temporally chaotic it makes
sense to introduce an effective {\em temperature\/}. In the
dimensionless units of the equation of motion (\ref{INT:eqm}) it is
the average kinetic energy in the frame co-moving with the center of
mass, i.e.,
\begin{equation}
  T\equiv\left\langle\!\!\left\langle\frac{(\dot x_j-v)^2}{2}
   \right\rangle\!\!\right\rangle,
  \label{STC:T}
\end{equation}
where $v$ is the average sliding velocity
\begin{equation}
  v=\langle\!\langle\dot x_j\rangle\!\rangle.
  \label{STC:v}
\end{equation}
In the previous paper we have derived a formula for the applied force
$F$ in terms of the first and second moments of the particle velocity
[Eq.~(\arabic{US:Fv2}) in Ref.~\onlinecite{str97}].  With the help of
this formula the temperature can be expressed in terms of the
applied force and the average sliding velocity:
\begin{equation}
  T=\left(\frac{F}{\gamma}-v\right)\frac{v}{2}.
  \label{STC:TF}
\end{equation}
Figure~\ref{f.vF.chaos}(b) shows the temperatures of the fluid-sliding
states of the velocity-force characteristic in
Fig.~\ref{f.vF.chaos}(a).

Even though the temperature of the solid-sliding states and the
stationary states is formally zero in accordance with (\ref{STC:TF}),
it does not make sense to call (\ref{STC:T}) a ``temperature'' in
regular, non-chaotic sliding states or stationary states. The
periodic and quasi-periodic domain-like states, for example,
investigated in the previous paper have also non-zero ``temperature''.

\begin{figure}
\epsfxsize=80mm\epsffile{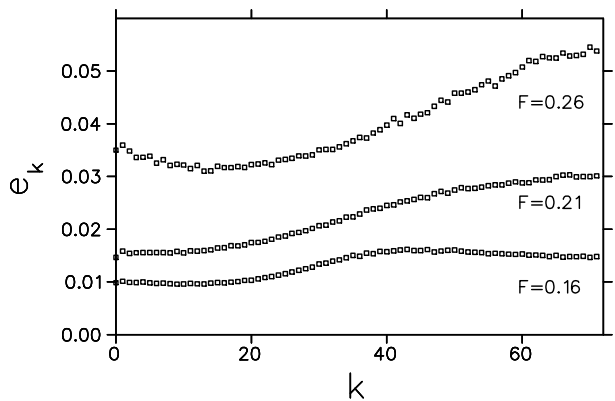}\vspace{2mm}
\caption[Phonon modes]{\protect\label{f.phmod}
The average kinetic energy $e_k$ of the phonon modes for three 
different values of the applied force $F$. The parameters are the
same as in Fig.~\ref{f.vF.chaos}.
}
\end{figure}

In the frame co-moving with the center of mass, the chain is shaken
by the washboard wave (i.e., the external potential) and moves in a
spatio-temporally chaotic way. The gaussian distributed velocities 
might suggest that the chain is in thermal equilibrium.  But is it
true?  This raises the following question of general interest: {\em Can we
replace the spatio-temporally chaotic chain by an equivalent system
which is in thermal equilibrium\/}? Or more general, is it possible to
describe the chaotic attractor of a weakly damped and strongly driven
Hamiltonian system with many degrees of freedom (infinitely many in
the thermodynamic limit) by an equivalent undriven and undamped
system?  As a consequence of a positive answer one would expect that
the equipartition theorem from thermodynamics holds, i.e.  the
ensemble averages of $q_j\partial H/\partial q_j$ and $p_j\partial
H/\partial p_j$ are independent of $j$ 
[$H(q_1,p_1,\ldots,q_j,p_j,\ldots)$ is the Hamilton function, the
$q_j$'s are the generalized coordinates, and the $p_j$'s are the
corresponding canonical momenta]. In numerical simulations one
usually replaces the ensemble average by the temporal average
assuming that the ergodicity hypothesis holds. An obvious candidate
for a test of the equipartition theorem would be the particle
momentum in the co-moving frame, i.e.,  $\dot x_j-v$. But this is not
a wise choice: because of symmetry the result has to independent of
$j$. A better choice is its spatial Fourier transform, i.e.,
\begin{equation}
  \hat{p}_k\equiv\frac{1}{N}\sum_{j=1}^N (\dot x_j-v)e^{2i\pi kj/N},
   \quad k=0,1,\ldots,N-1.
  \label{STC:pk}
\end{equation}
That is, we want to check whether the average kinetic energy
\begin{equation}
  e_k\equiv\lim_{\tau\to\infty}\frac{1}{\tau}\int_0^\tau
   |\hat{p}_k(t)|^2dt
  \label{STC:ek}
\end{equation}
of the phonon modes is equipartitioned or not. Figure~\ref{f.phmod}
shows that the equipartition theorem is not fulfilled \cite{rem0}. This is a
clear signature for the fact that the {\em fluid-sliding state is a
state far away from thermal equilibrium\/}. Therefore it is not
possible to develop a theory for this state based on equilibrium
thermodynamic. The violation of the equipartition theorem is
equivalent with non-zero velocity correlations for $j\neq 0$ because
$e_k$ is the modulus of the Fourier transform of $C_j$.

\subsection{\protect\label{SST}The transition between fluid-sliding 
state and kink-dominated sliding state}

\begin{figure}
\epsfxsize=80mm\epsffile{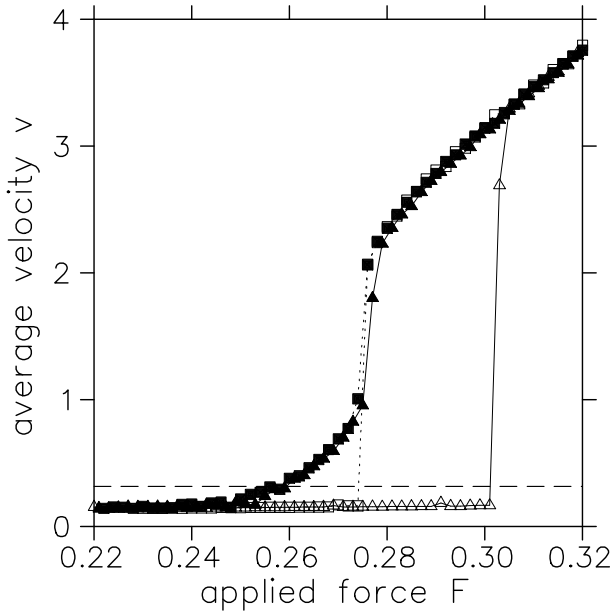}\vspace{2mm}
\caption[Velocity-force characteristic 2]{\protect\label{f.vF.chao2}
The transition between the fluid-sliding state and the kink-dominated
sliding state. The
velocity-force characteristics for $N=500$ (filled symbols) and 
$N=200$ (open symbols) are shown.  Squares and dotted lines
(triangles and solid lines) indicate decreasing (increasing) applied
force $F$.  The rate $|\dot F|$ is always $10^{-7}$ except for
$N=500$ and $F\in (0.26,0.29)$ where it is $10^{-8}$.  The dashed
line indicates the sound velocity which is equal to $a$.  The
parameters are $a=\pi/10$, $b=2$, and $\gamma=0.05$.
}
\end{figure}

When $a/2\pi$ approaches an integer value the velocity-force
characteristic of the fluid-sliding state develops a relatively sharp
transition step at a characteristic value of the applied force $F$.
An example for $a/2\pi=1/20$ is shown in
figure~\ref{f.vF.chao2}. For long enough chains no hysteresis is
observable. For small chains we get bistability between different
types of sliding states. Similar results has been found in a generalized FK
model by Braun {\em et al.\/} \cite{bra97a,bra97b,pal97}. 
The aim of this section is to answer the
following obvious questions: What is the nature of the different
sliding states? Why does the bistability depends on $N$? Can we
understand this transition and where does it occur?

\begin{figure}
\epsfxsize=80mm\epsffile{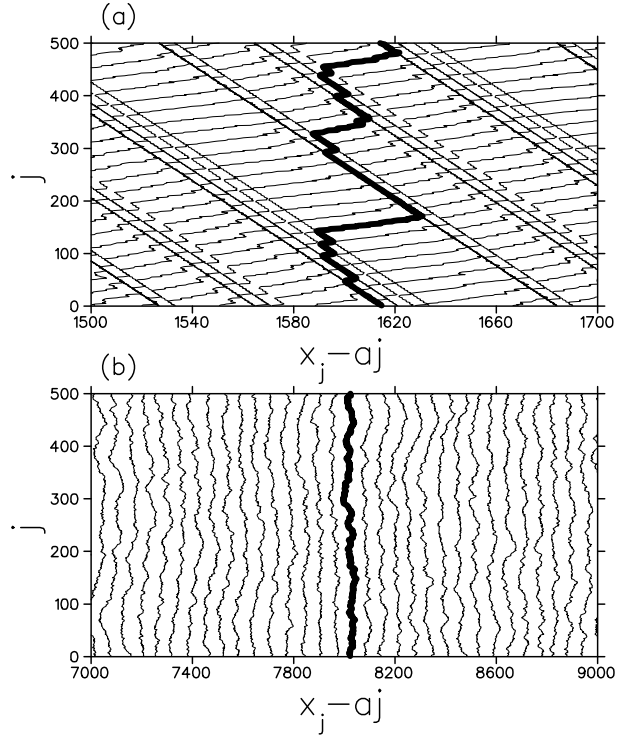}\vspace{2mm}
\caption[The transition]{\protect\label{f.sst}
The dynamics of the kink-dominated sliding state (a) and the 
fluid-sliding state (b). Several snapshots are
shown taken at equidistant time steps (a) $\delta t=2\pi/v$ and (b)
$\delta t=20\pi/v$. In each case a particular snapshot is 
highlighted. The parameters are (a) $F=0.2$ and (b) $F=0.3$ and 
$N=500$, $M=25$, (i.e., $a=\pi/10$), $b=2$, and $\gamma=0.05$.  }
\end{figure}

First we take a more detailed look at the dynamics below and above
the transition (see Fig.~\ref{f.sst}).  The motion far below the
threshold is almost regular. It corresponds to one of the
multi-domain states we have discussed in the previous paper. There
are two domain types, a stationary one with $a=0$ [it is responsible
for the tilted lines in Fig.~\ref{f.sst}(a)] and a sliding one. Often
the sliding domains are so small that they are actually
$2\pi$-kinks, and larger sliding domains can be interpreted as
clusters of $2\pi$-kinks \cite{bra97a}. That is, multi-domain states
like the example of Fig.~\ref{f.sst}(a) are nonuniform distributions
of $2\pi$-kinks. Thus, we call this state the {\em kink-dominated 
sliding state\/}. Note, that there are $M$ kinks but no antikink.

It is well-known that kinks (and antikinks) cannot travel faster than
the sound velocity (which is equal to one in our case). Each kink or
antikink needs therefore at least $N$ time units to travel through
the whole chain. After that time a chain with $M$ kinks will be
shifted by $2\pi M$. Therefore, the average sliding velocity of a
state like the one of Fig.~\ref{f.sst}(a) has to be less than $2\pi
M/N=a$.  Fig.~\ref{f.vF.chao2} shows that it is actually much below
the sound velocity. 

\begin{figure}
\epsfxsize=80mm\epsffile{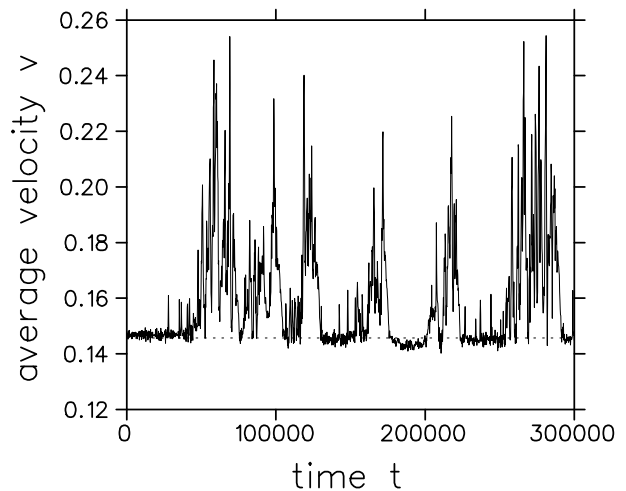}\vspace{2mm}
\caption[Velocity fluctuations]{\protect\label{f.fluct}
The average sliding velocity $v$ (averaged over intervals of 200
time units) as function of time. The velocity $v_0$ of the quiescent
phase is denoted by a dotted line.  The parameters are $N=500$,
$M=25$, (i.e., $a=\pi/10$), $b=2$, $\gamma=0.05$, and $F=0.24$.
}
\end{figure}

If the transition point is approached from below, the behavior
depends on whether the chain is long or short. For long chains with
many kinks, the average sliding velocity $v$ starts to increase with
$F$ faster and faster. Later on the increase slows down.  We define
the transition point $F_{\rm FKT}$ as the value of $F$ where the
slope of $v(F)$ has a maximum. After the transition point the system
is in a fluid-sliding state [see
Fig.~\ref{f.sst}(b)]. All kinks
(and antikinks) have disappeared, and the system is completely
spatio-temporally chaotic. A short chain with only a few kinks still
stays in the kink-dominated regime beyond $F_{\rm FKT}$. Eventually, it
jumps to the fluid-sliding state or to the solid-sliding
state (see Fig.~\ref{f.vF.chao2}). A hysteresis occurs, and the chain
goes back to the kink-dominated state at $F\approx F_{\rm FKT}$.

\begin{figure}
\epsfxsize=80mm\epsffile{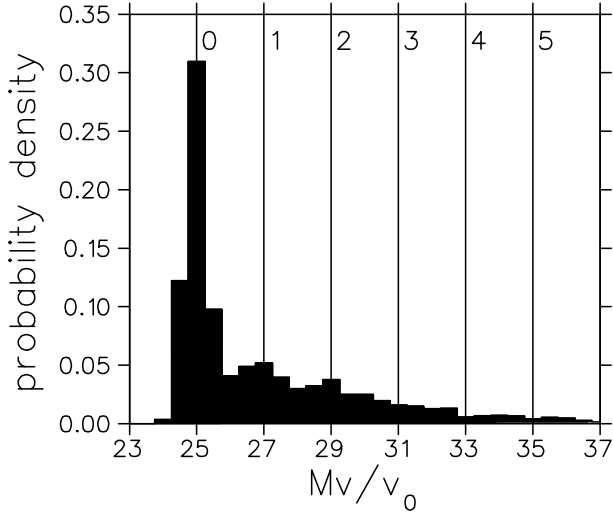}\vspace{2mm}
\caption[Velocity distribution]{\protect\label{f.fluctd}
The velocity distribution in the kink-dominated regime near the 
transition point. The statistics is obtained from 5000 samples. 
Each sample is the sliding velocity $v$
averaged over 200 time units. The velocity of the quiescent phase is
given by $v_0\approx 0.1457$. The numbered vertical lines denote the
number $N_p$ of kink-antikink pairs.  The parameters are the same as
in Fig.~\ref{f.fluct}.
}
\end{figure}

At the transition point the sliding velocity strongly fluctuates.
These fluctuations set in already much below the 
transition point. They lead to a larger value of the average sliding
velocity compared to the value for small chains. A typical example is
shown in figure~\ref{f.fluct}. One sees bursts of activity above a
level given by the value of $v$ for small chains (see
Fig.~\ref{f.vF.chao2}). A detailed look onto the dynamics of the chain
reveals that an increase of $v$ is caused by the {\em production of
kink-antikink pairs\/} \cite{bra97a,bra97b,pal97}. These pairs usually appear
behind a $2\pi$-kink cluster. This may be the reason why for small
chains the kink-dominated states survive beyond the transition point
because the probability for a $2\pi$-kink cluster is too small. Each
kink and antikink contributes to the sliding velocity of the chain.
That is, $v$ is given by
\begin{equation}
  v=2\pi\frac{M+2N_p}{N}c_k,
  \label{SST:v}
\end{equation}
where $c_k$ is the velocity of the kinks and antikinks and $N_p$ is
the number of kink-antikink pairs. In the quiescent state without
kink-antikink pairs, the average sliding velocity is given by
$v_0=2\pi Mc_k/N$. Thus the ratio $v/(v_0/M)$ gives directly the
total number of kinks and antikinks (i.e., $M+2N_p$).
Figure~\ref{f.fluctd} depicts the distribution of the sliding velocity
shown in Fig.~\ref{f.fluct}. In addition to the large peak at the
quiescent state one sees clearly small peaks for $N_p=1$ and $2$.

For $a$ approaching zero (or any other integer multiple of $2\pi$)
the average sliding velocity of the kink-dominated sliding state also
approaches zero. It disappears at $a=0$. Thus, the system goes from
the fluid-sliding state directly to a stationary state when $F$ is
decreased below the transition point $F_{\rm FKT}$. If the value of
$b$ is not too large [i.e., $b={\cal O}(1)$], the stationary state
will be the ground state \cite{rem1}.  The transition is often
accompanied by transients where a few kink-antikink pairs survive for
a considerably long time. But in all simulations these pairs
eventually disappeared.

We found that the transition point $F_{\rm FKT}$ is nearly
independent of $a$ (as long as $a/2\pi$ is near an integer value).
Table~\ref{t.fsst} shows that $F_{\rm FKT}$ increases with $b$ weaker
than linearly. 

In order to understand the transition between the fluid-sliding state 
and the kink-dominated sliding state we compare
the FK model with a simpler model, namely, one particle in a tilted
spatially periodic potential plus additive white noise. This model
was studied in details by Risken and Vollmer \cite{ris84,vol80}. The
single particle and the noise correspond to the center of mass of the
FK model and the chaotic motion of the internal degrees of freedom,
respectively. Of course the noise is neither additive nor white. Its
strength depends on the state. It is obvious that the solid-sliding
state and the stationary state of the FK model correspond to the
running state and the locked state of the simple model in the absence
of noise.  We suggest that the fluid-sliding state and the kink-dominated
state of the FK model correspond also to the running and locked
states of the simple model but now with noise. Risken and Vollmer
showed that the bistability between the running state and the locked
state disappears even for infinitesimally weak noise. There is a
well-defined transition point $F_2$ which is smeared out for
increasing noise strength. For $\gamma\ll \sqrt{b}$ it is given by
\begin{equation}
  F_2=3.3576\gamma\sqrt{b}.
  \label{SST:F2}
\end{equation}
Thus we expect $F_{\rm FKT}\approx F_2$.
Table~\ref{t.fsst} shows that this is indeed the case especially for
small values of $b$. 

\begin{table}
\caption[Transition point]{\protect\label{t.fsst}The 
transition point $F_{\rm FKT}$ between fluid-sliding states and 
kink-dominated sliding states and the transition
point $F_2$ between the running state and the locked state of single
particle in a periodic potential under the influence of weak noise.
The parameters are $N=500$, $M=0$, and $\gamma=0.05$.}
%\squeezetable
\begin{tabular}{ccc}
$b$&$F_{\rm FKT}$&$F_2$\\
\\ \hline
0.25 & 0.096 & 0.0840\\
0.5 & 0.130 & 0.1187 \\
1 & 0.187 & 0.1679 \\
2 & 0.276 & 0.2374 \\
3 & 0.362 & 0.2908 \\
4 & 0.444 & 0.3358 \\
5 & 0.519 & 0.3754 \\
\end{tabular}
\end{table}

\section{\protect\label{MF}Nonequilibrium freezing and melting}

The aim of this section is to take a closer look at the hysteresis
loop between stationary states and the fluid-sliding state (see
Fig.~\ref{f.vF.chaos}). There are two different types of transitions.
A depinning-pinning transition where the (chaotically) sliding chain
turns into a stationary one. It is a kind of nonequilibrium freezing
and, as in usual first-order phase transitions, the chain can be
``supercooled'' below the transition point $F_{\rm DP}$. That is, for
$F$ slightly below $F_{\rm DP}$ the chain does not freeze
immediately. It takes a while until a critical nucleus has appeared.
The second transition is the pinning-depinning transition. The
transition point $F_{\rm PD}$ depends on the stationary state and
therefore on the history of the system.  In an actual experiment
where one sweeps through the hysteresis loop at a finite rate, one
will therefore not get well-defined transitions points but more or
less broad distributions. An example of such a (numerical) experiment
is shown in the inset of Fig.~\ref{f.vF.chaos}(a). It is typical that
the distribution of $F_{\rm DP}$ is narrower than the distribution of
$F_{\rm PD}$. For a quasistatic sweep the distribution of $F_{\rm
DP}$ becomes sharp, whereas the width of the distribution of $F_{\rm
PD}$ is nearly independent on the sweeping rate.

From equilibrium thermodynamics it is well-known that melting and
freezing occur at the same temperature and the transition is of
first-order. In our case far from thermal equilibrium the situation
is different. In the overdamped limit the pinning-depinning
transition point $F_{\rm PD}$ is identical with the depinning-pinning
transition $F_{\rm DP}$ but the transition is of second order
\cite{cop88,flo96}. In the underdamped case we have bistability
between sliding states and stationary states, i.e., $F_{\rm
DP}<F_{\rm PD}$. 

\begin{figure}
\epsfxsize=80mm\epsffile{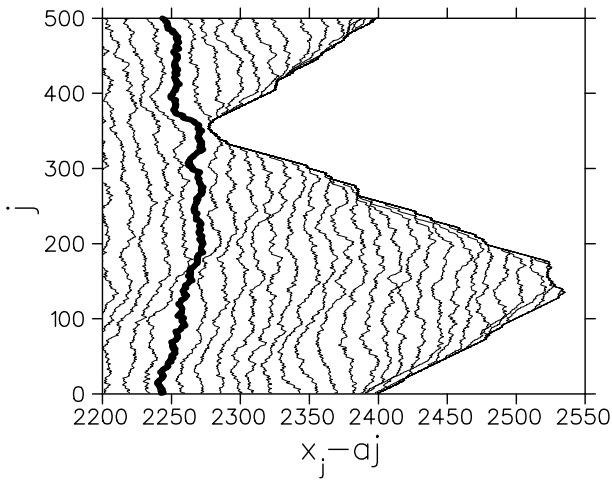}\vspace{2mm}
\caption[Nucleation]{\protect\label{f.nuc}
Nucleation of a stationary state in the depinning-pinning transitions.
The time step between two snapshots is $\delta t=4\pi/v$. The
snapshot just before nucleation is highlighted.
The parameters are $N=500$, $M=160$, 
$b=2$, $\gamma=0.05$, and $F=0.097$.
}
\end{figure}

A further characteristic feature of nonequilibrium melting and
freezing is the change of the effective temperature [see e.g.
Fig.~\ref{f.vF.chaos}(b)]. This is a general property which is also
found in similar models with nonzero temperature of the environment
\cite{bra97a,per93,gra96}. That is, the fluid-sliding state has an
effective temperature which is {\em independent\/}  on the
temperature of the environment as long as the latter one is
considerably less than the former one. The effective temperature is a
measure of the enhanced energy flow due to the excitation of phonons.

\subsection{\protect\label{DPT}Freezing: 
The depinning-pinning transition}

We have studied in detail the depinning-pinning transition from a
fluid-sliding state to a stationary state. We have chosen a value of
$a$ where the fluid-sliding state is completely spatio-temporally
chaotic. Like in an ordinary first-order phase transition at thermal
equilibrium the transition is caused by a {\em nucleation process\/}.
That is, a small portion of the chain becomes stationary, and the
fronts between this stationary nucleus and the sliding chain
propagate into the chain. An example of such a nucleation process is
shown in figure~\ref{f.nuc}. One sees that states appear behind the
fronts which can be characterized by an average particle distance $a$
(or density $1/a$) which is roughly constant. Note that the values of
$a$ behind both fronts have to be different. This can be understood
by the following argument. In the previous paper we have seen that
because of the conservation of the number of particles the velocity
of a front traveling from one particle to the next is given by
$c=(v_1-v_2)/(a_2-a_1)$, where the average particle distance and the
average sliding velocity on both sides of the front are given by
$a_{1/2}$ and $v_{1/2}$, respectively. The chaotic sliding state is
characterized by $a_1=2\pi M/N$ and $v_1>0$. For the stationary state
$v_2=0$ holds. In order to have fronts traveling in opposite
directions (see Fig.~\ref{f.nuc}) the average particle distances of
the stationary states have to be different (one larger than $a$ and
one smaller than $a$). We found that they are always the nearest
integer multiples of $\pi$, that is, for $0<a<\pi$ they are zero and
$\pi$. From this consideration it is clear that after the
depinning-pinning transition the system cannot be in the ground state
\cite{rem1} which is characterized by a single domain with a uniform
$a$.  Instead the chain will be in a stationary {\em two-domain
state\/} which is not quite perfect because each domain contains
defects at low density. 

A similar depinning-pinning transition was found in a seemingly
completely different model, namely, the quenched Kadar-Parisi-Zhang
equation with negative nonlinearity \cite{jeo96}. It is a nonlinear
diffusion equation which models the surface growth in disordered
media. Here the dissipation is strong and the randomness of the
pinning landscape is important contrary to the weakly damped FK
model. Nevertheless the pinning-depinning transition is of first
order and the state which freezes out of the sliding state is a
two-domain state, as in the FK model (compare Fig.~2 of
Ref.~\onlinecite{jeo96} with Fig.~\ref{f.nuc}).

The duration of freezing $t_D$ is the sum of the nucleation time
$t_N$ and the growth time $t_G$. The nucleation time is the time that
evolves until a critical nucleus appears. A critical nucleus is a
nucleus which is large enough to grow into the fluid-sliding state.
The nucleation time $t_N$ will be a Poisson distribution if the
probability for the appearance of a critical nucleus in a short
time-step is small, i.e.,
\begin{equation}
 \rho(t_N)=\theta^{-1}e^{-t_N/\theta}.
  \label{DPT:rho}
\end{equation}
It is well-known that the average and the standard deviation of a
Poisson-distributed value are identical, i.e., $\langle
t_N\rangle=\Delta t_N=\theta$. For systems much larger than the
critical nucleus, the probability for nucleation increases linearly
with the system size, i.e., $\theta\sim 1/N$. 

The time $t_G$ for a critical nucleus to grow up to the stationary
state does not fluctuate as strongly as $t_N$ because it is roughly
given by the system size divided by the sum of front velocities. Thus
$\langle t_G\rangle\sim N$.

\begin{figure}
\epsfxsize=80mm\epsffile{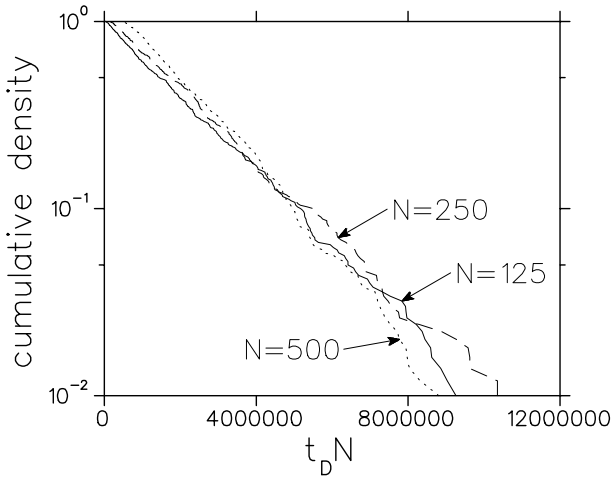}\vspace{2mm}
\caption[TD distribution]{\protect\label{f.poi} Cumulative density of
the duration times $t_D$. Each curve is obtained from 500 numerical
nucleation experiments. The parameters are $a=16\pi/25$, $b=2$,
$\gamma=0.05$, and $F=0.097$.
}
\end{figure}

Figure~\ref{f.poi} shows that the nucleation process is indeed
characterized by a Poisson distribution with $\theta\sim 1/N$. From
fits of the curves shown in Fig.~\ref{f.poi} we found $\theta
N=(2.1\pm 0.1)\cdot 10^6$. One can also see a shift of the
distribution for increasing $N$ to larger values of $t_D$. This
reflects the fact that $\langle t_G\rangle$ increases with $N$.

For values of $a$ between 2.45 and 3.09 (and $b=2$, $\gamma=0.05$,
$N=500$, and $F\approx F_{\rm FKT}$)  we found that the fluid-sliding
state changes its character. A domain appears where the particles are
stationary (with two particles per potential well).  This domain
which is surrounded by spatio-temporal chaos has a constant size and
travels through the chain with a constant velocity. The transition
from this so-called {\em traffic-jam state\/} \cite{bra97a} 
to a stationary one occurs also via nucleation. The
critical nucleus seems to appear always at the back of the stationary
domain.  It reverses the propagation direction of the front. That is,
the chaotic state stops traveling into the stationary one.  Instead a
front propagates into the chaotic state leaving behind a stationary
state different from the already existing one.  Thus the result is
again a stationary two-domain state where one domain already exists
at least partially before the transition. The nucleation probability
is roughly independent of the system size $N$ because the nucleation
site is predetermined.

\begin{figure}
\epsfxsize=80mm\epsffile{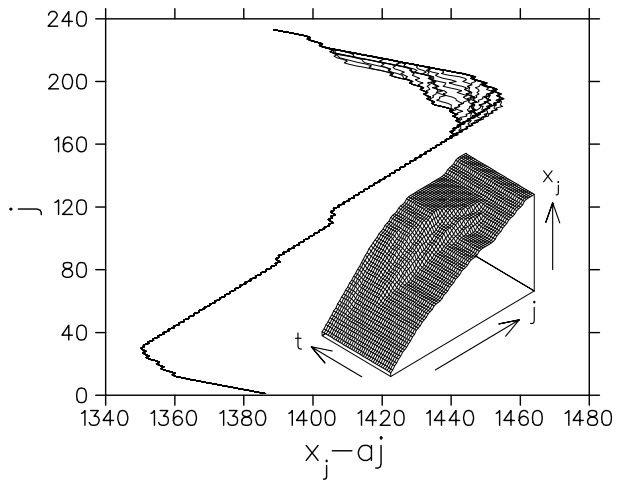}\vspace{2mm}
\caption[Micro slip]{\protect\label{f.mslip}
An example of a micro-slip. Snapshots at an interval of $10$ 
time units are shown. The inset shows $x_j$ for $j\in[160,233]$.
The parameters are
$N=233$, $M=89$, $b=2$, $\gamma=0.05$, and $F$ from $0.11554$
until $0.115575$ at a rate of $10^{-7}$.
}
\end{figure}

\subsection{\protect\label{PDT}Melting: Pinning-depinning transition}

The transition from a stationary chain to a sliding one is the
pinning-depinning transition. The transition point $F_{\rm PD}$
depends on the stationary state. It corresponds to a saddle-node
bifurcation where the particular stationary state annihilates with an
unstable stationary state. In the overdamped limit $F_{\rm PD}$ is
uniquely defined by the saddle-node bifurcation of the last stable
stationary state which is usually the state which develops
adiabatically out of the ground state for $F=0$. The disappearances
of the other stationary states in a saddle-node bifurcation lead only
to more or less local rearrangements of the chain. We call such a
rearrangements a {\em micro-slip\/}. A micro-slip changes the center
of mass of the chain but does not lead to sliding. An example is
shown in figure~\ref{f.mslip}.  This behavior and the statistics of 
micro-slips has been studied mainly in models with random external
potential \cite{mid91,pla91}.

In the underdamped regime the energy gained from a local
rearrangement is not dissipated immediately. This may lead to an
avalanche which turns the whole system into a sliding state. Whether
a saddle-node bifurcation leads to a micro-slip or to a transition to
sliding depends on the stationary state and on the damping constant
$\gamma$. For each stationary state one can presumably find a
critical value $\gamma_c$ which distinguishes both cases: For
$\gamma>\gamma_c$ we get a micro-slip, otherwise a transition to
sliding. In our simulations we found only a few (not more than three)
micro-slips for $\gamma=0.05$ and $b=2$. The pinning-depinning
transition point depends strongly on the history of the system
because the actual value of $F_{\rm PD}$ depends on the stationary
state.  Sweeping several times through the pinning-depinning
hysteresis loops, we find a broad distribution of $F_{\rm PD}$ even
for very small sweep rates. 

\section{\protect\label{CON}Summary and concluding remarks}

In this paper we have shown that in the {\em weakly\/} damped and
strongly driven FK model spatio-temporal chaos appears. This chaotic
state, called {\em fluid-sliding state\/}, has an effective
temperature which reflects the fact that phonons are excited.  This
has to be contrasted by the solid-sliding state where no phonons are
excited \cite{rem2}. The excitation of phonons opens up additional
channels of dissipation. Thus friction in the fluid-sliding state is
larger than in the solid-sliding state. 

The velocity-force characteristic shows a pronounced transition if
the average particle distance $a$ is near but not identical to an
integer multiple of the period of the external potential. It is a
transition between the kink-dominated sliding state and the
fluid-sliding state.  Below the transition point $F_{\rm FKT}$
sliding is caused by the propagation of $2\pi$ kinks. Approaching the
transition point from below leads to an increasing production of
kink-antikink pairs. Above the transition point all kinks and
antikinks disappear and the chain is in the fluid-sliding state with
an average sliding velocity which depends only weakly on the average
particle density $1/a$. In the kink-dominated regime the average
sliding velocity is proportional to $a\,{\rm mod}\,2\pi$. The
transition point $F_{\rm FKT}$ is roughly independent of $a$. For
small values of $b$ it is very well approximated by the transition
point between the running and locked solution of an ($N=1$)-FK model
with infinitesimally small additive white noise [i.e.
Eq.~(\ref{SST:F2})]. This raises the question whether the dynamics
can be reduced to a center-of-mass motion plus some nontrivial (e.g.
colored, state-dependent) noise term.

The nonequilibrium freezing (i.e., the transition from sliding to
stationarity) of the fluid-sliding state is a nucleation process. The
resulting stationary state has two domains of different particle
densities. 

The nonequilibrium melting (i.e., the transition from a stationary
state to sliding) of such a stationary state corresponds to a
saddle-node bifurcation where the stationary and, of course, the
stable state annihilates with its unstable counterpart. In the case
of nonzero environmental temperature the transition occurs a bit
earlier because thermal activation overcomes the barrier which
decreases to zero at the saddle-node bifurcation.  Not all
saddle-node bifurcations lead to melting. They may lead to a local
rearrangement of the particle configuration called micro-slip. The
pinning-depinning transition point depends on the history because
each stationary state has another bifurcation point. 

Our results are qualitatively very similar to results of similar
models. Persson studied a two-dimensional Lennard-Jones liquid on a
corrugated potential with square symmetry at a nonzero temperature
\cite{per93}.  He varied the temperature between $1/3$ and $1/2$ of
the melting temperature (the thermal energy was roughly a tenth of
activation energy for single particle diffusion). He obtained
velocity-force characteristics and effective temperature plots
similar to Fig~\ref{f.vF.chaos}. Because of the relatively high
temperature of the environment he did not find hysteresis between the
solid-sliding state and the fluid-sliding state. As in the FK model,
the transition from the fluid-sliding state to a stationary state is
also a nucleation process. Persson found that it occurs when the
effective temperature is equal to the melting temperature at thermal
equilibrium. Because the velocity distribution of the fluid-sliding
state was Gaussian, he argued that the fluid-sliding state is in a
kind of thermal equilibrium. Therefore it has to freeze below the
melting temperature. But as we have seen in Sec.~\ref{STC}, Gaussian 
distributed velocities do not imply a quasi-thermal equilibrium.  It
would be interesting to look whether in Persson's model the
equipartition theorem of the phonon modes is fulfilled or not. This
is clearly a better test on thermal equilibrium.

Granato {\em et al.\/} studied a two-dimensional Frenkel-Kontorova
model at nonzero temperatures \cite{gra96}. The external potential is
corrugated only in the direction of the applied force. Also the
particles can move only in this direction. They did the simulations
at a temperature which corresponds to $1/4$ of the activation energy
and which is roughly $1/5$ of the melting temperature. They also
found a behavior like in Fig.~\ref{f.vF.chaos} again without a
hysteresis between the solid-sliding state and fluid-sliding state.
They also confirmed the observation of Persson that the transition
from sliding to stationarity occurs at the melting temperature.

Braun {\em et al.\/} investigated a generalized FK model that is 
quite similar to the model
studied by Persson \cite{bra97a,bra97b,pal97}. The main difference is
that the substrate potential is anharmonic and the interaction
potential is an exponential repulsion. They did the simulations
mainly for two different temperatures which correspond to $10^{-3}$
and $1/20$ of the activation energy for single particle diffusion.
They report results for one-dimensional chains of atoms as well as
for two-dimensional layers. There seems to be no qualitative
differences between one and two dimension. Again velocity-force
characteristics and effective temperature plots are similar to our
results. They found a hysteresis between solid sliding and fluid
sliding. Its width decreases with the environmental temperature and
disappears eventually (at a temperature which is roughly $1/5$ of the
activation barrier for hopping of uncoupled single particles).  Near
fully commensurate particle densities they found a transition
(denoted by $F_{\rm pair}$) which is similar to the transition from
the kink-dominated sliding regime to the fluid-sliding regime of the
FK model.  Because their chain was short (i.e., $N=105$) this
transition shows hysteresis for very low temperatures.  Instead of a
fully chaotic fluid-sliding regime they always found two-domain
states with alternatively running and locked particles (traffic-jam
regime). They also measured the transition point $F_{\rm pair}$ as
function of the damping constant \cite{pal97}. It increases roughly 
linear with the damping constant but the authors seemed not to be
aware that (\ref{SST:F2}) is again a remarkable good approximation 
(i.e. errors are less than 10\%) for $F_{\rm pair}$.

\acknowledgments

We thank H. Thomas for critical reading of the manuscript. We also
acknowledge the possibility to do simulations on the NEC SX-3 and
SX-4 at the Centro Svizzero di Calcolo Scientifico at Manno,
Switzerland. This work was supported by the Swiss National Science
Foundation.


\begin{references}

\bibitem{rub89}
M.A. Rubio, C.A. Edwards, A. Dougherty, and J.P. Gollub, Phys. Rev. 
Lett. {\bf 63}, 1635 (1989).

\bibitem{he92}
S. He, G. Kahamanda, and P.-Z. Wong, Phys. Rev. Lett. {\bf 69}, 3731 
(1992).

\bibitem{bla94}
G. Blatter, M.V. Feigel'man, V.B. Geshkenbein, A.I. Larkin, and 
V.M. Vinokur, Rev. Mod. Phys. {\bf 66}, 1125 (1994).

\bibitem{gru88}
G. Gr\"uner, Rev. Mod. Phys. {\bf 60}, 1129 (1988).

\bibitem{ris84}
H. Risken, {\it The Fokker-Planck Equation}, (Springer, Berlin, 1984).

\bibitem{nat92}
T. Natterman, S. Stepanow, L.H. Tang, and H. Leschom, J. Phys. II 
(France) {\bf 2}, 1483 (1992).

\bibitem{nar93}
O. Narayan and D. Fisher, Phys. Rev. B {\bf 48}, 7030 (1993).

\bibitem{kon38}
T. Kontorova and J. Frenkel, Z. Phys. Sowjetunion {\bf 13}, 1 (1938).

\bibitem{wat96}
S. Watanabe, H.S.J. van der Zant, S.H. Strogatz, and T.P. Orlando,
Physica D {\bf 97}, 429 (1996).

\bibitem{bra97a}
O.M. Braun, T. Dauxois, M.V. Paliy, and M. Peyrard, Phys. Rev. Lett. 
{\bf 78}, 1295 (1997); Phys. Rev. E {\bf 55}, 3598 (1997).

\bibitem{str97}
T. Strunz and F.J. Elmer, previous paper.

\bibitem{per93}
B.N.J. Persson, Phys. Rev. B {\bf 48}, 18140 (1993).

\bibitem{gra96}
E. Granato, M.~R. Baldan, and S.~C. Ying, in {\em The physics of 
sliding friction\/}, B.~N.~J. Persson and E. Tosatti (eds.), 
(Kluwer Academic Publishers, Dordrecht, 1996).

\bibitem{bra97b}
O.M. Braun, A.R. Bishop, and J. R\"oder, Phys. Rev. Lett. {\bf 79},
3692 (1997).

\bibitem{cro93}
M.C. Cross and P.C. Hohenberg, Rev. Mod. Phys. {\bf 65}, 851 (1993).

\bibitem{eck85}
J.P. Eckmann and D. Ruelle, Rev. Mod. Phys. {\bf 57}, 617 (1985).

\bibitem{rem0}
By the way, we checked in microcanonical simulations of the thermal
equlibrium (i.e., no damping, $\gamma=0$, no driving, $F=0$), 
that for the same values of $b$ and $T$ as in Fig.~\ref{f.phmod} 
the kinetic energy $e_k$ of the phonon modes are indeed equipartitioned.

\bibitem{pal97}
M. Paliy, O. Braun, T. Dauxois, and B. Hu, Phys. Rev. E {\bf 56},
4025 (1997).

\bibitem{rem1}
More precisely: The state which develops adiabatically out of the
ground state of the undriven system (i.e., $F=0$).

\bibitem{vol80}
H.D. Vollmer and H. Risken, Z. Phys. B {\bf 37}, 343 (1980).

\bibitem{cop88}
S.N. Coppersmith and D.S. Fisher, Phys. Rev. A {\bf 38}, 6338 (1988).

\bibitem{flo96}
L.M. Flor\'\i a and J.J. Mazo, Adv. Phys. {\bf 45}, 505 (1996).

\bibitem{jeo96}
H. Jeong, B, Kahng, and D. Kim, Phys. Rev. Lett. {\bf 77}, 5094 
(1996).

\bibitem{mid91}
A.A. Middleton and D.S. Fisher, Phys. Rev. Lett. {\bf 66}, 92 (1991).

\bibitem{pla91}
O. Pla and F. Nori, Phys. Rev. Lett. {\bf 67}, 919 (1991).

\bibitem{rem2}
There is one exception, namely, the 
phonon with the same wave number as the washboard wave. But its
amplitude is very small (see Sec.~III.B in Ref.~\onlinecite{str97}).

\end{references}
\end{document}